\documentclass[final,1p,times]{elsarticle} 
\usepackage{graphicx} 
\usepackage{amssymb} 
\usepackage{amsthm} 
\usepackage{lineno} 
\usepackage{hyperref}

\def\lsim{\mbox{~{\protect\raisebox{0.4ex}{$<$}}\hspace{-1.1em}
	{\protect\raisebox{-0.6ex}{$\sim$}}~}}
\def\Eq#1{Eq.~(\ref{#1})}

\def\Fig#1{Fig.~\ref{#1}}
\def\Sect#1{Section~\ref{#1}}
\def\Ref#1{Ref.\cite{#1}}
\def\Refs#1{Refs.\cite{#1}}

\def\st{\begin{equation}}
\def\stp{\end{equation}}
\def\bg{\begin{eqnarray}}
\def\nd{\end{eqnarray}}
\def\llangle{\left\langle}
\def\rrangle{\right\rangle}

\def\x{{\bf x}}

\def\pr{{\mathcal P}}

\def\nc{{\, ,}}
\def\np{{\, .}}

\journal{Nuclear Physics A} 
\begin{document} 

\begin{frontmatter} 


\title{
A Summary of Bulk Dynamics from Quark Matter 2009}

\author{Derek Teaney$^{a,b}$}

\address[a]{ Department of Physics and Astronomy,
Stony Brook University,
Stony Brook, NY 11794-3800, USA}
\address[b]{ 
RIKEN-BNL Research Center, 
Building 510A, Physics Department,
Brookhaven National Laboratory, 
Upton, NY 11973-5000, USA} 

\begin{abstract} 
I review the recent progress in measuring elliptic flow  in heavy ion collisions.
These measurements show clearly how hydrodynamics starts to develop as the 
system size is increased from peripheral to central collisions.
During this transition, the momentum range described by hydrodynamics increases
as the system progresses from a kinetic to a hydrodynamic regime. Many of the
systematic deviations from ideal hydrodynamics  are reproduced effortlessly
once the shear viscosity is included.  In order to extract the shear viscosity
from the data, kinetic theory can be used to determine which aspects of the 
elliptic flow reflect the details of the microscopic interactions, and which
aspects reflect the underlying transport coefficients. I also review the identified
hadron elliptic flow and the predictions of hydrodynamics for the LHC.
\end{abstract} 

\end{frontmatter} 



\section{Overview}

Perhaps the most important result from the Relativistic Heavy Ion 
Collider is the observation of strong elliptic flow \cite{Voloshin:2008dg,TeaneyQGP4}. Elliptic 
flow is an asymmetry of particle production with respect to 
the reaction plane and has been measured as 
a function of transverse momentum, rapidity, and particle type. The interpretation
of the observed flow which has been adopted by the heavy
ion community is that  the elliptic flow is the hydrodynamic
response to the collision geometry.
Implicit in this interpretation of the observed flow is that the time scale 
for momentum relaxation near the QCD phase transition is 
of order the quantum time scale
\st
   \tau_R \sim  \frac{\hbar}{\pi T} \, .
\label{tauq}
\stp
This estimate for the relaxation time is best expressed in terms of the
shear viscosity  to entropy ratio. For instance, in the viscous Bjorken 
model the energy density  at time $\tau_o$ evolves as 
\st
 \frac{de}{d\tau} = - \frac{e + \pr}{\tau_o} + \frac{4}{3} \frac{\eta}{\tau^2_o}\, ,
\stp
where $e$ is the energy density, $\pr$ is the pressure, and $\eta$ is the shear viscosity \cite{Danielewicz:1984ww}.
Comparing the size of the viscous term to the ideal term, we conclude that
hydrodynamics will provide a good description of the observed flow when
\st
   \frac{\eta}{(e + \pr)\tau_o} \ll 1 \, .
\stp
Using $e + \pr = sT$, and an estimate for  the temperature and $\tau_o$, this
criterion reads
\st
    0.2 \left( \frac{\eta/s}{0.3} \right) 
\left(\frac{1\, {\rm fm} }{\tau_o} \right) \left( \frac{300\,\mbox{MeV}}{T_o} \right)  \ll 1 \np
\label{etabys}
\stp
From this estimate we see that hydrodynamics will begin to be a good
approximation  for $\eta/s \lsim 0.3$ or so.   To reiterate, implicit in the
hydrodynamic interpretation of the flow results is a strong conclusion about
the transport coefficients of QCD. 

Many of the systematic trends
seen in the elliptic flow data {\it do} support the hydrodynamic 
interpretation of the observed flow.  
For example, as a function 
of centrality and transverse momentum, the measured elliptic
flow deviates from ideal hydrodynamics  in a way characteristic of
viscosity. 
These experimental trends have been fully
clarified only recently  and the  experimental analysis is now rather
sophisticated. These developments  were reported on at the Quark Matter conference  and are  reviewed in \Sect{experiment}.  The systematics of the recent 
flow measurements give confidence in the overall picture of the hydrodynamic expansion.

Although these experimental trends support the notion of a
hydrodynamic response, there are several puzzling patterns in  the elliptic flow
data.  For example,  new data on the elliptic flow of the $\phi$ meson and the $\Omega^{-}$ baryon
are reviewed in \Sect{experiment}.  The
differences in the measured flow 
between mesons and baryons is generally explained with the
coalescence model, which enjoys considerable phenomenological success. 
However, the coalescence model is theoretically  unsatisfactory and is difficult to realize in a dynamical model.
The seemingly simple coalescence trends
seen in the elliptic flow data must be understood before the shear viscosity
and other transport properties can be reliably extracted from the heavy ion
data.

In addition to experimental progress, there 
has been substantial theoretical progress in classifying the form of viscous
corrections, both with viscous hydrodynamics and with kinetic theory.  
A brief summary of some of the developments discussed at Quark Matter 2009
is  presented in \Sect{Theory}.  Most of these ideas presented in this summary are reviewed more completely in \Ref{TeaneyQGP4},  which was written shortly 
after  the Quark Matter conference.  Some sections from this longer review have 
been copied for this brief summary.

\section{Measurements}
\label{experiment}

One of the best ways to test the hydrodynamic interpretation is to systematically observe how the response changes from small systems to large systems. 
Experimentally this can be accomplished by colliding small nuclei
such as CuCu and selecting peripheral collisions. Unfortunately measuring elliptic flow
in these smaller systems is rather difficult, and reliable, precise results 
have become  available only recently.   

The difficulty in measuring flow 
in small systems  stems from fluctuations.
Especially in peripheral AuAu and CuCu collisions,  there are fluctuations in the initial eccentricity of the 
participants.
Thus rather than using the continuum approximation 
to categorize the geometry, it is better to 
implement a Monte-Carlo Glauber calculation 
and estimate the  eccentricity using the ``participant plane eccentricity". 
This event by event eccentricity is denoted $\epsilon_{PP}$ in the literature. 
Clearly the experimental goal is to extract the {\it response} 
coefficient $C$ relating
the elliptic flow to the eccentricity on an event by event basis
\st
    v_2  = C \epsilon_{PP} \np
\stp

\begin{figure}
\centering
\includegraphics[width=0.75\textwidth]{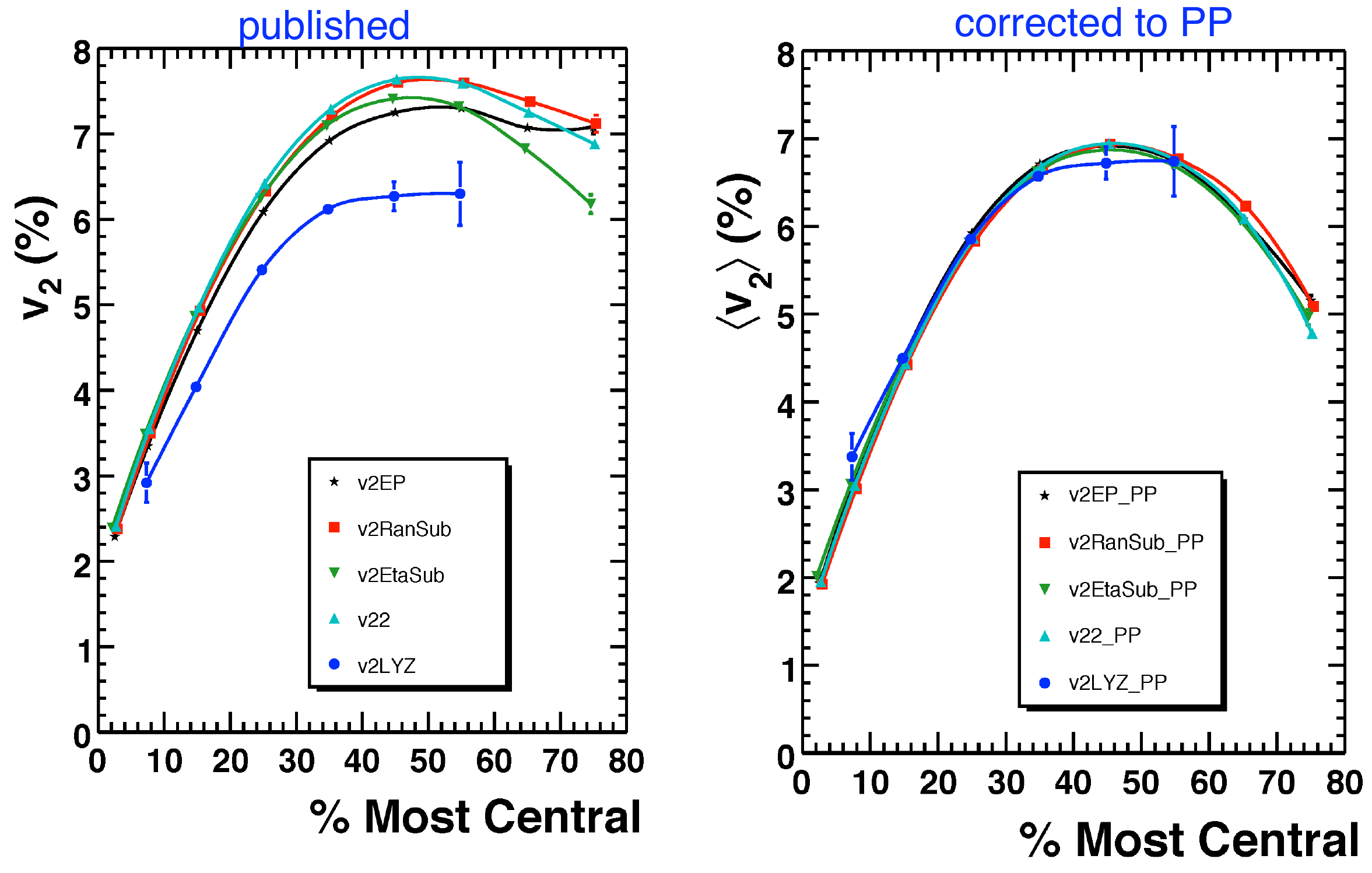}
\caption[]{(a) The elliptic flow measured 
by different methods \cite{Ollitrault:2009ie}. The methods measure different quantities and should agree
only when fluctuations are neglected. (b) 
The average elliptic flow $\llangle v_2 \rrangle$ measured by the different
methods. To determine $\llangle v_2 \rrangle$ for each method,  non-flow was estimated using measured $pp$ data and fluctuations were estimated with a geometric model.
 }
\label{fig:art}
\end{figure}

If the flow methods measured $\llangle v_2 \rrangle$,  then we 
could  simply divide the measured flow to determine the response coefficient, $C=\llangle v_2 \rrangle / \llangle \epsilon_{PP} \rrangle$.
The PHOBOS collaboration deciphered
the  confusing CuCu data by recognizing the need for $\epsilon_{PP}$ and
following this procedure\cite{Alver:2006wh}.
However, it was generally realized (see in particular. \Ref{Miller:2003kd}) that the elliptic flow methods
 do not measure precisely $\llangle v_2\rrangle$. Some methods (such as two particle correlations $v_2\left\{2\right\}$)
are sensitive  to $\sqrt{\llangle v^2_2 \rrangle}$, while other methods (such 
as the event   plane method $v_2\left\{EP\right\}$) measure  
something closer to $\llangle v_2 \rrangle$. What 
precisely the event plane method measures depends on 
the reaction plane resolution in a known way\cite{Ollitrault:2009ie}.
So just dividing the measured flow by the average participant eccentricity is not entirely 
correct.
The appropriate quantity to divide by depends on the method 
\cite{Miller:2003kd,Bhalerao:2006tp,Voloshin:2007pc}.
In a Gaussian approximation for the eccentricity fluctuations 
this can  be worked out analytically. For instance, the two particle correlation
elliptic flow  $v_{2}\left\{2\right\}$ (which measures
$\sqrt{\llangle v_2^2 \rrangle}$),  should be divided by 
$
   \sqrt{ \llangle \epsilon_{PP}^2 \rrangle } \np
$
With a complete understanding of what each method measures, \Ref{Ollitrault:2009ie} 
was able to make a simple model for the 
fluctuations and non-flow and show that $\llangle v_2 \rrangle$ measured by
the different methods are compatible to high precision.  
This is illustrated in \Fig{fig:art}. The stunning precision of the recent flow 
measurements will be very useful in understanding how hydrodynamics  begins to 
work in peripheral collisions. 
This work should be extended to the CuCu system where non-Gaussian
fluctuations are stronger and ultimately corroborate the 
PHOBOS  analysis \cite{Alver:2006wh,Alver:2008zza}. This is a worthwhile goal 
because it will clarify fully the transition into the hydrodynamic regime \cite{Drescher:2007cd,Gombeaud_this_work}.

\begin{figure}
\centering
\includegraphics[width=0.45\textwidth]{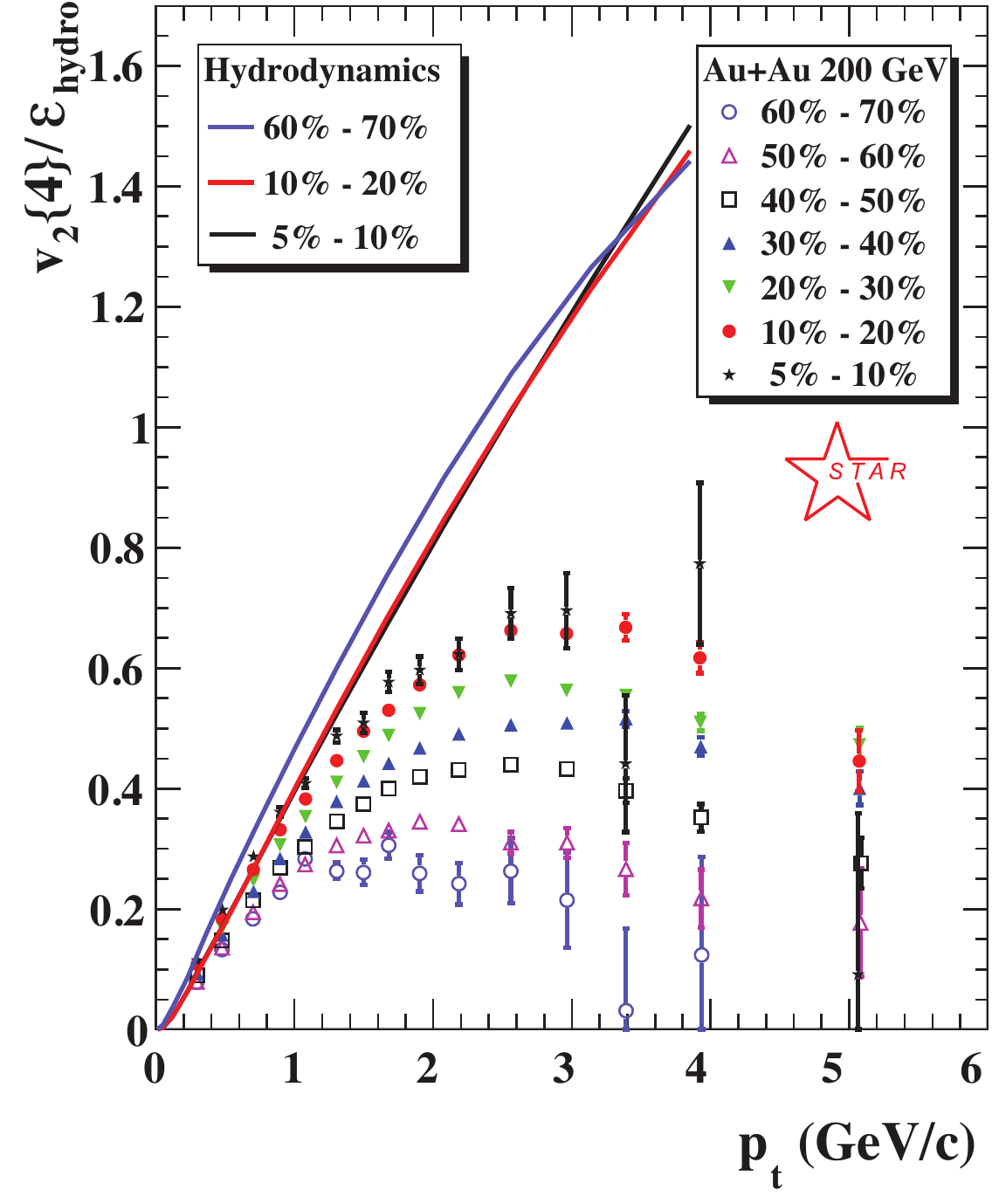}
\caption[]{
Elliptic flow $v_2(p_T)$ as measured by the STAR collaboration \protect \cite{STARV2,STAR:2008ed} for different centralities.  
The measured elliptic flow has been divided by the eccentricity.
The curves are ideal hydrodynamic calculations based
on \protect\Refs{Huovinen:2006jp,Huovinen:2001cy} rather than the viscous hydrodynamics discussed in much of this summary. 
}
\label{fig:raimond} 
\end{figure}

The scaled elliptic flow $v_2/\epsilon$
measures the response of the medium to the initial geometry  and
can be used to determine the regime of validity of hydrodynamics. 
Additional information about the shear viscosity can be  gleaned from the transverse momentum dependence of the observed 
elliptic flow.
\Fig{fig:raimond} shows $v_2(p_T)/\epsilon$ as a
function of centrality, 0-5\% being the most central and 60-70\% being the most
peripheral. Examining this figure we see a gradual transition  from a weak to a
strong dynamic response with growing system size.  The
interpretation adopted here is that this  change is a consequence of
a system transitioning from a kinetic to a hydrodynamic regime. 

There are several theoretical curves based upon calculations
of ideal hydrodynamics\cite{Huovinen:2001cy,Kolb:2000fh} 
which for $p_T <  1\,{\rm GeV}$ approximately
reproduce the observed elliptic flow in the most central collisions.
Since ideal hydrodynamics 
is scale invariant (for a scale invariant equation of state) the expectation
is that the response $v_2/\epsilon$ of this theory should 
be independent of system size or centrality. This reasoning is borne out 
by the more elaborate hydrodynamic calculations shown in the figure.
On the other hand, the data show a gradual transition as a function of
increasing centrality, rising towards the ideal hydrodynamic calculations  in
a systematic way.
These trends are captured by models with a finite mean free path\cite{Drescher:2007uh}. 

The data show other trends as a function of centrality. In more 
central collisions the linearly rising trend, which resembles
the ideal hydrodynamic calculations,
extends to larger and larger transverse momentum.
Viscous corrections to ideal hydrodynamics grow as 
\st
   \left(\frac{p_T}{T}\right)^2 \frac{\ell_{\rm \scriptscriptstyle mfp}}{L} \nc
\stp
where $L$ is a characteristic length scale. Thus 
these viscous corrections restrict the applicable momentum range in hydrodynamics \cite{Teaney:2003kp}.
In more central collisions, where $\ell_{\rm \scriptscriptstyle mfp}/L$ 
is smaller,  the transverse momentum range described by 
hydrodynamics extends to increasingly large $p_T$.  These qualitative
trends are reproduced by the more involved viscous calculations \cite{TeaneyQGP4}.



\begin{figure}
\centering
\includegraphics[height=2.3in]{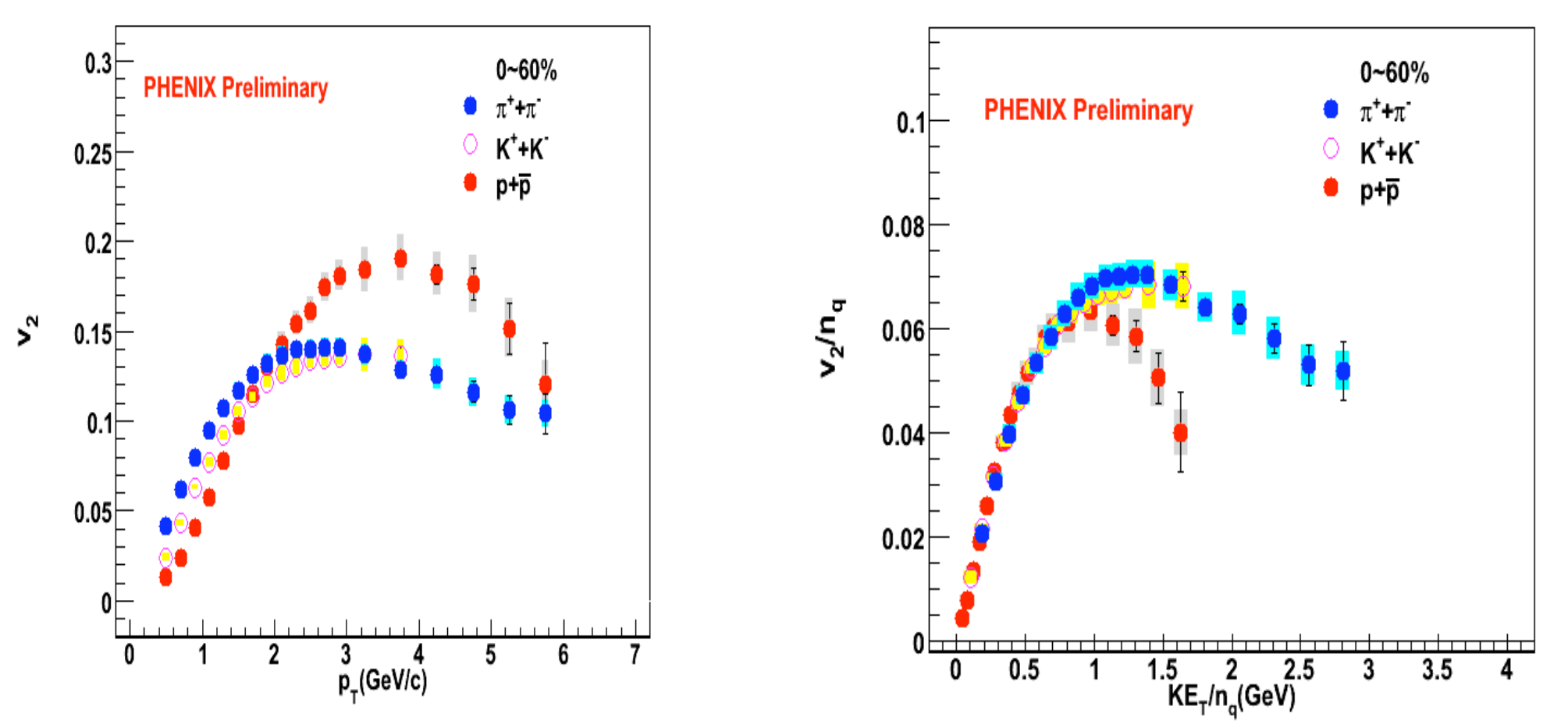}
\caption[]{(a) The elliptic flow as a function of transverse momentum 
for identified particles as measured by the 
PHENIX collaboration \cite{phenix_data}. (b) The elliptic flow of identified hadrons rescaled 
according to the quark coalescence model and plotted as a function  of $KE_T = \sqrt{p_T^2 + M^2} - M$.  
}
\label{fig:roygroup}
\end{figure}

While many of the trends seen in \Fig{fig:art} and \Fig{fig:raimond} 
are reproduced and understood with viscous hydrodynamics, there are 
additional trends in the elliptic flow data which are only partially understood.
For instance  \Fig{fig:roygroup}(a) shows the elliptic flow of identified particles
$\pi,K,p$. At low momenta the separation amongst the different 
particle species is well reproduced by hydrodynamics.  However as the 
momentum is increased the  proton elliptic flow equals, and then exceeds, the 
pion elliptic flow. These systematic
trends are seen in all collision systems and centralities.  The prevailing 
explanation  is that constituent quarks coalesce to form hadrons at the 
phase boundary. 
This ansatz is supported by the observation that if the hadron momentum and $v_2$ is divided by the  valence quark content  all of  the $v_2$ of the different
hadron species lie along a single curve. This is illustrated in 
\Fig{fig:roygroup}(b) which is plotted as a function of $KE_T = \sqrt{p_T^2 + M^2} - M$ rather than transverse momentum to capture the hydrodynamic behavior
at small momentum.  Although constituent quark scaling works rather well, the theoretical support for quark coalescence is small since it is difficult 
to realize a coalescence mechanism  in a dynamical simulation.  It nevertheless remains to
find an alternative picture for the observed different flows of  mesons and
baryons. At the Quark Matter conference the elliptic flow of identified particles was measured accurately out to rather large transverse momenta. At sufficiently large momenta the data deviate from a universal coalescence curve providing new insight into the hadronization dynamics in this region.

\begin{figure}
\centering
\includegraphics[width=0.75\textwidth]{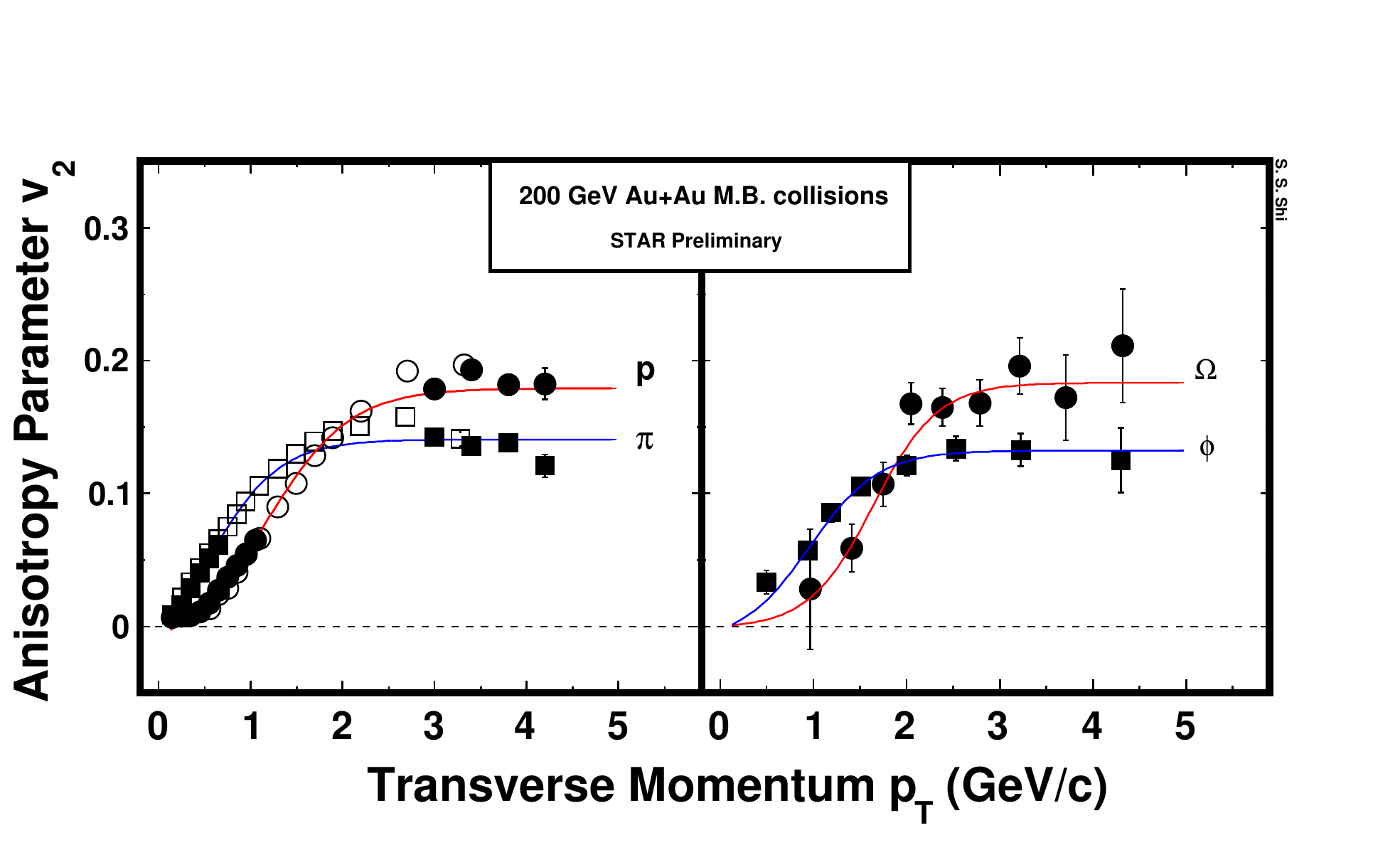}
\caption{A comparison of the elliptic flow of pions and protons 
to the elliptic flow  of the multi-strange $\phi$ and $\Omega^{-}$ \cite{Shi:2009jg}. }
\label{fig:staromega}
\end{figure}

To conclude this section, we turn to \Fig{fig:staromega} which compares the elliptic flow
protons and pions to the 
flow of the multi-strange hadrons $\Omega^{-}$ and $\phi$. 
The important point
is that the $\Omega^{-}$ is nearly twice as heavy as the proton
and more importantly, does not have a strong resonant interaction
analogous to the $\Delta$. For these reasons the hadronic relaxation time
of the $\Omega^{-}$ is expected to be much longer than the duration 
of the heavy ion event. Nevertheless the $\Omega^-$ shows nearly
the same elliptic flow as the protons.  This provides fairly convincing
evidence that the majority of the elliptic flow develops during 
a deconfined phase which 
hadronizes to produce a flowing $\Omega^{-}$ baryon. 

\section{Modeling Elliptic Flow with Hydrodynamics and Kinetic Theory}
\label{Theory}

Many of the trends seen in the data (with the possible exception of quark
coalescence) are reproduced by hydrodynamics and kinetic theory. Generally the
kinetic theory estimates for the shear viscosity are consistent with the
estimates from viscous hydrodynamics. Specifically unless $\eta/s \lsim 0.3$ it
is impossible to reproduce the observed elliptic flow.  Given that the
transport time scales extracted from the heavy ion data are close to the
quantum time scale given in \Eq{tauq},  it is clear that the microscopic
details of kinetic models cannot be trusted.  Nevertheless these shortcomings of
the microscopic theory are  unimportant in the hydrodynamic regime. In the
hydrodynamic regime  the only properties that determine the evolution of the
system are the equation of state, $\pr(e)$, and the shear viscosity and bulk
viscosities, $\eta(e)$ and $\zeta(e)$.  In the sense that kinetic theory
provides a reasonable guess as to how the surface to volume ratio influences
the forward evolution, these models can be used to estimate  the shear
viscosity, and the estimate  may be  more reliable than the hydrodynamic models.
More importantly,  by comparing the results of different microscopic models one
can determine which features of the heavy ion data are universal ($i.e.$ only
depend on $\eta(e),\pr(e)$ and $\zeta(e)$ ).  

\begin{figure}
\centering
\begin{minipage}[c]{0.55\textwidth}
\includegraphics[width=0.9\textwidth]{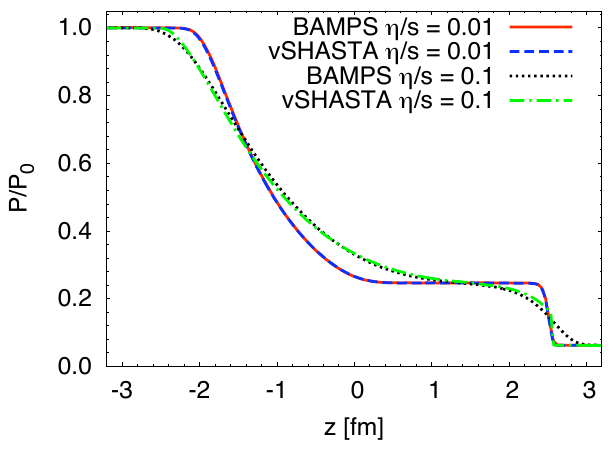}
\end{minipage}
\begin{minipage}[c]{0.32\textwidth}
\includegraphics[width=\textwidth]{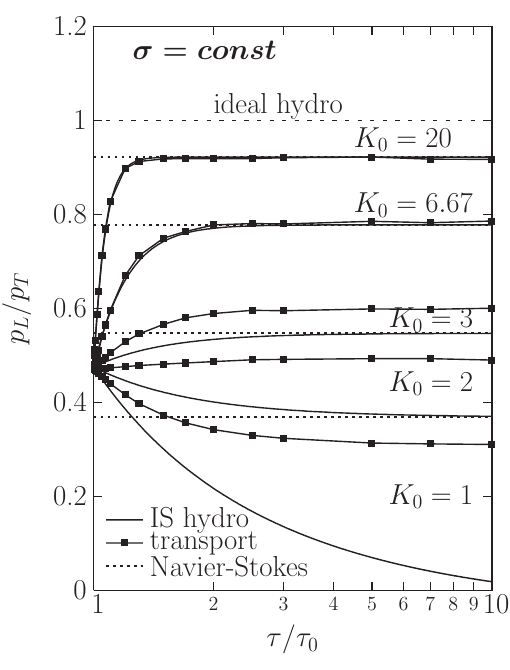}
\end{minipage}
\caption{(a) The pressure relative to the initial pressure
for shock tube initial conditions. The three curves are 
for the BAMPS parton cascade model, viscous hydrodynamics (vSHASTA), and ideal 
hydrodynamics
\protect\cite{Bouras:2009nn}. 
(b) The ratio of longitudinal and transverse pressures in  
kinetic theory, Israel Stewart hydrodynamics, and the Navier-Stokes 
equations, for  a Bjorken expansion 
of a massless ideal gas with constant cross section. 
The simulations are compared as a function of Knudsen number, $K\equiv \sigma/\pi R^2 \, dN/dy$ \protect\cite{Huovinen:2008te}. 
}
\label{Bouras}
\end{figure}

There were a number of promising efforts to reproduce hydrodynamic results from 
kinetic theory reported at the conference. First there was an effort to reproduce hydrodynamic shocks with
kinetic theory by the Frankfurt group. \Fig{Bouras}(a) 
shows a kinetic theory simulation of the shock tube problem.  Clearly the BAMPS
code is capable of reproducing the correct hydrodynamic limit in detail. The
deviations of the BAMPS code from ideal hydro are beautifully reproduced by
viscous hydro. This gives a great deal of confidence in the BAMPS code and in
the viscous hydrodynamic code vSHASTA. Additional results from the BAMPS
simulation of elliptic flow were presented in the poster session.  

A similar approach to the hydrodynamic limit was   reported by Huovinnen and Molnar \cite{Huovinen:2008te}, and Gombeaud \cite{Gombeaud_this_work}. In particular \Fig{Bouras}(b) shows a simulation of the MPC code which also makes a direct comparison with viscous hydrodynamics for the Bjorken expansion of 
an ideal massless gas with constant cross section. As the Knudsen number
$K\equiv \sigma /\pi R^2 \,dN/dy$  is increased, the simulation approaches
Israel-Stewart hydro and ultimately the Navier-Stokes limit.
Given that the kinetic code and the viscous hydrodynamic simulations agree reasonably, the MPC code can be used to reliably extract the shear viscosity from the heavy ion data.

When viscous hydrodynamics is extended to a second order there are additional
relaxation times (e.g. $\tau_\pi$) beyond the shear viscosity, $\eta(e)$.  Just as transport
models should be approximately independent of the details of the microscopic
interactions, results from viscous hydrodynamics should be approximately
independent of the precise way in which the second order terms are implemented. 
This is indeed the case \cite{Luzum:2008cw,Dusling:2007gi,Song:2007ux}.
Additional results from viscous hydrodynamics will be discussed 
more completely by P. Romatschke in this volume \cite{Romatschke_this_vol}.

\section{Outlook}
\begin{figure}
\centering
\includegraphics[scale=0.41]{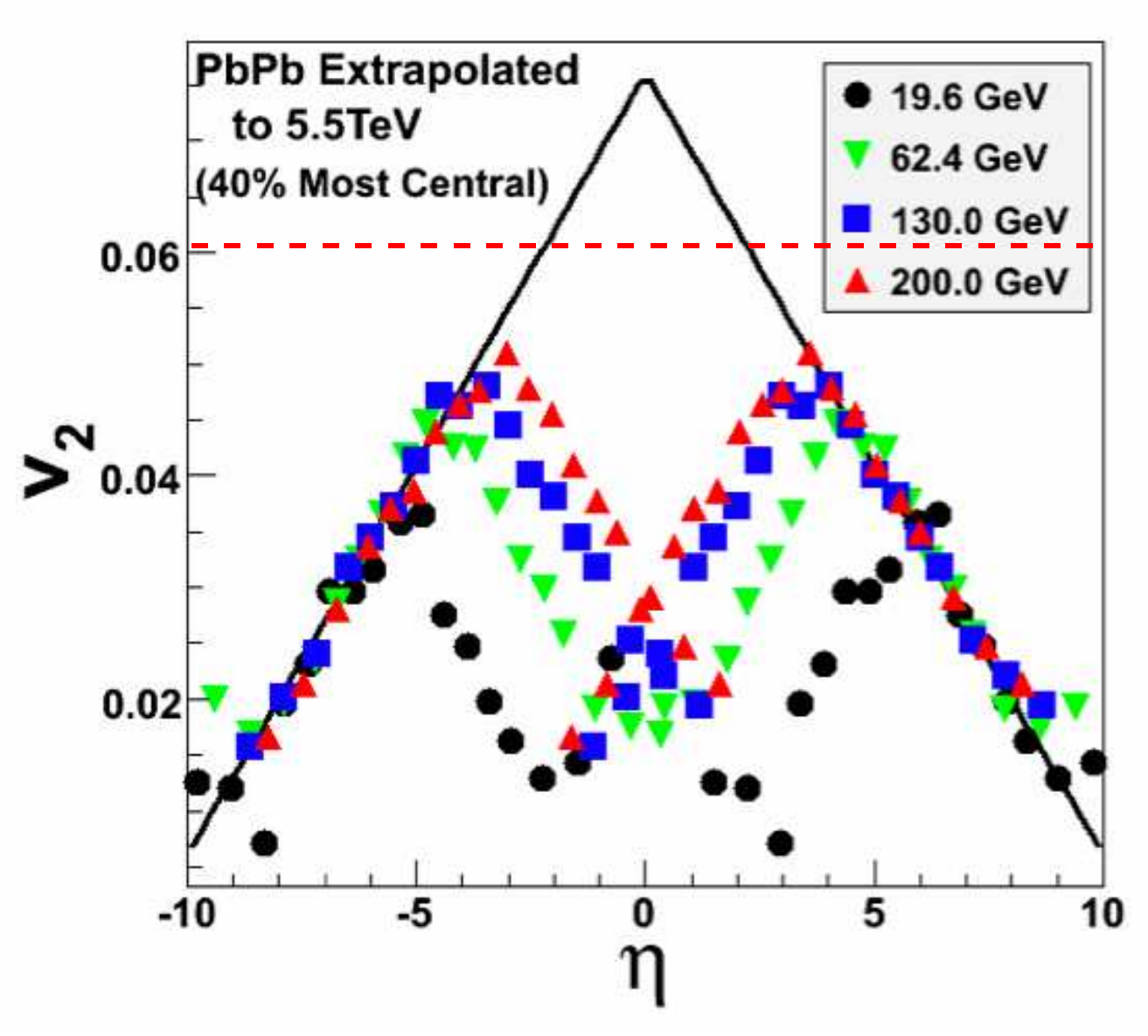}
\caption[]{Figure presented by W. Busza providing an estimate for  
the elliptic flow at the LHC by extrapolating trend lines from lower 
energy to higher energy (solid black lines) \cite{Busza_this_vol}.  The dashed red lines shows 
an estimate based on ideal hydrodynamics \protect\cite{Teaney:2001av}.}
\label{fig:lhc}
\end{figure}

Clearly there is a strong convergence between  kinetic 
and  hydrodynamic simulations of heavy ion reactions.  
These simulations reproduce many trends observed in increasingly
precise measurements of elliptic flow. 
This convergence strongly suggests that the hydrodynamic interpretation 
of the observed flow is correct. One of the striking tests of hydrodynamic 
predictions is the saturation of elliptic flow at high energy. From 
RHIC to the LHC, hydrodynamics
predicts an increase  in the flow which is significantly less than a 
naive extrapolation from lower energies. This is illustrated in \Fig{fig:lhc} 
and will be one of the first tests of the hydrodynamic paradigm at the LHC.

\section*{Acknowledgments} 
This summary was a result of extensive discussions at the Quark Matter meeting  with 
Peter Arnold, Aihong Tang, Kevin Dusling, Raimond Snellings, Paul Romatschke, Peter Petreczky,  Mikko Laine, Steffan Bass, Vincenzo Greco, Denes Molnar, Fuqiang Wang, Sergei Voloshin. 
This
work was supported in part by   
by the U.S. Department of Energy under an OJI
grant DE-FG02-08ER41540 and as a RIKEN and Sloan Fellow.

\end{document}